\documentclass[usenatbib]{mn2e}
\usepackage{aas_macros}
\usepackage{graphicx}

\def\sun{\odot}

\title{Halo Concentration and the Dark Matter Power Spectrum}

\author[K. Huffenberger \& U. Seljak]{Kevin M. Huffenberger \& Uro\v s Seljak\\
Department of Physics, Jadwin Hall\\ Princeton University, Princeton, NJ 08544}

\date{January 2003}

\begin{document}

\maketitle

\begin{abstract}
We explore the connection between halo concentration and the dark matter 
power spectrum using the halo model. We fit halo model parameters
to non-linear power spectra over a large range of cosmological models.
We find that the non-linear evolution 
of the power spectrum generically 
prefers the concentration  
at non-linear mass scale to decrease with the effective slope of the linear 
power spectrum, in agreement with the
direct analysis of the halo structure in different cosmological models. 
Using these analyses, we compute the 
predictions for non-linear power spectrum beyond the current resolution 
of $N$-body simulations. We find that the halo model predictions are 
generically below the analytical non-linear models, suggesting that the 
latter may overestimate the amount of power on small scales. 

\end{abstract}

\section{Introduction}
In recent years, computational $N$-body simulations have provided important insights into the non-linear formation of structure by the dark matter.  
 \citet{1997ApJ...490..493N} (NFW) observed a `universal' density profile for dark matter haloes, suggesting that the profiles depend only on the 
mass of the halo, with relatively little scatter between them. 
They also  noticed that the halo concentration parameter, describing the degree 
of profile steepness (defined more precisely below)
 decreases with increasing halo mass, though the exact relation depends on the cosmology at hand.  
Several formulae have been suggested for this relation \citep{1997ApJ...490..493N,2001MNRAS.321..559B,2001ApJ...554..114E} which rely on fitting some mechanism to haloes culled from $N$-body simulations.
These studies found that the concentration dependence on the power 
spectrum is caused primarily by the slope of the linear power spectrum. 
Both this and the mass dependence of concentration 
can be explained by the epoch of the halo formation, since less massive 
haloes form at higher redshift when the universe was denser and this leads 
to a more concentrated profile.

Another relation which appears to be universal is that of the 
mass function \citep{2001MNRAS.321..372J}. The distribution of halo masses in $N$-body 
simulations appears to be universal in the sense that when 
mass is related to the fluctuation amplitude of the linear power 
spectrum, the mass function has a universal form. This relation has 
also been explored in detail in recent work \citep{1999MNRAS.308..119S,2001MNRAS.321..372J,2001A&A...367...27W,2002astro.ph..7185W}, defining precisely what 
is meant by the mass and confirming the universality of the mass 
function for a large set of cosmological models. 

While the above approaches have focused on individual haloes in the simulations
by counting them and exploring their structure, for many cosmological 
applications all that we need is the dark matter power spectrum. 
This can be computed from $N$-body simulations as well, 
but computational limitations 
prevent one from exploring a large range of cosmological models or 
from achieving a high resolution on small scales. 
For this reason many fitting formulae have been developed with increasing 
accuracy over the years \citep{1991ApJ...374L...1H,1995MNRAS.276L..25J,1996MNRAS.280L..19P,2002astro.ph..7664S}, although these are still significantly limited 
on small scales by the dynamical range of simulations. 

The close connection between the two descriptions has been highlighted 
recently with the revival of the halo model \citep{2000MNRAS.318..203S,2000MNRAS.318.1144P,2000ApJ...543..503M,2001ApJ...546...20S}. 
For a few cosmological models, it was shown that with an appropriate choice of concentration mass 
dependence and mass function one can use the halo model to accurately predict the non-linear dark matter power spectrum. 
One of the questions we explore in this 
paper is whether this agreement can be extended to a wider range of models
and whether the trends seen in the analysis of individual haloes are 
confirmed also with the power spectrum analysis. There are several reasons
why this is of interest: 
since the selection of individual haloes  
is somewhat subjective it is possible that those left out may be 
systematically different. For example, they could be less relaxed, more 
recently formed
and so less
concentrated. An opposite effect is that because the power 
spectrum is a pair-weighted statistic, if there is a scatter in 
the mass-concentration relation then pair weighting increases 
the mean concentration relative to the simple particle 
weighting. 

On the other hand, if the analysis of individual haloes is in agreement 
with the power spectrum analysis over the range where both are reliable, 
then the same analysis can be extended 
to smaller scales which are not resolved by cosmological $N$-body simulations 
that compute the power spectrum. The reason for this is that the 
resolution scale for the halo structure can be extended significantly 
by resimulating the representative regions of the haloes with a higher 
resolution simulation zoomed on that halo. On the other hand, a power 
spectrum calculation requires a large simulation volume, so that the 
largest scales are in the linear regime, which then implies that the 
mass and force resolution cannot be as high as in the simulations which 
focus on single haloes. 

This paper is structured in the following way. 
We summarize the halo model and give relevant background information in \S\ref{sec:Background}.  In \S\ref{sec:Method}, we give our method and results.  In \S\ref{sec:Comparison}, we compare our results to other models  and in \S\ref{sec:Conclusions} we summarize our conclusions.

\section{Background} \label{sec:Background}

\subsection{The Halo Model} \label{subsec:halomodel}
We begin by reviewing the formalism of 
the halo model to compute 
the non-linear dark matter power spectra, following the notation in  
\citet{2000MNRAS.318..203S}. The haloes are characterized by their mass $M$, 
density profile $\rho(r,M)$, number density $n(M)$ and bias $b(M)$. 
$N$-body simulations suggest a family of density profiles:
\begin{equation}
\rho(r)={\rho_s \over (r/r_s)^{-\alpha}(1+r/r_s)^{3+\alpha}},
\label{eqn:rho(r)}
\end{equation}
where $\rho_s$ is a characteristic density, $r_s$ is the radius where the profile has an effective power law index of $-2$, and $-1.5<\alpha<-1$ \citep{1997ApJ...490..493N,1998ApJ...499L...5M}.  Here we use $\alpha = -1$, since power 
spectra are not sensitive to the inner parts of the halo.  

As is conventional, we re-parametrize $\rho_s$ and $r_s$ in equation (\ref{eqn:rho(r)}) in terms of a halo mass and concentration.  We define the virial radius $r_{\rm vir}$ of a halo to be the radius of a sphere with some characteristic mean density, discussed below.  Insisting that the mass contained inside $r_{\rm vir}$ is $M$ fixes $\rho_s$ for a given $r_s$.  We introduce the concentration parameter, the ratio $c \equiv r_{\rm vir}/r_s$.  
There are several definitions of virial radius used in the literature. 
In this paper, the virial radius is defined as the radius of a halo-centreed sphere which has a mean density 180 times the mean density of the universe.  
This may be denoted $r_{180\Omega}$.  Other authors set $r_{\rm vir}$ to the radius inside which the mean density is 200 times the {\em critical} density ($r_{200}$), a measure independent of $\Omega$ \citep[for example]{1997ApJ...490..493N}.  Still others use a spherical collapse model to estimate the radius of a virialized halo ($r_{\Delta}$).   
These lead to different concentrations and masses.  For $\Omega=0.3$ and $\Lambda=0.7$, $r_{200} < r_\Delta < r_{180\Omega}$, so for the same halo with 
$c_{180\Omega}=10$ one has $c_{200} \sim 6$, $c_\Delta\sim 8$, 
$M_{200}/M_{180\Omega} \sim 0.72$ and $M_\Delta/M_{180\Omega} \sim 0.87$.  
Assuming the NFW halo profile it is straightforward to translate values for concentrations and masses between these conventions.
We choose $c_{180\Omega}$ as it permits the use of the universal mass function \citep{2002astro.ph..7185W}.  To avoid cumbersome notation, in this paper we refer to $c_{180\Omega}$ as $c$ and $M_{180\Omega}$ as $M$.

For the mass dependence of concentration we choose a power law parametrization
\begin{equation}
c = c_0 \left({M/ M_*}\right)^\beta , \label{eqn:c(M)}
\end{equation} where $c_0$ and $\beta$ are free parameters, and $M_*$ is the non-linear mass scale, 
which we will define shortly. This is of course not the most general parametrization, but as we
will show below it suffices for the current dynamical range. 

The halo number density is written in terms of the multiplicity function via 
\begin{equation}
{n(M)}\, dM={\bar{\rho} \over M}f(\nu)\, d\nu,
\end{equation}
where $\bar{\rho}$ is the mean matter density of the universe and mass is written in terms of peak height
\begin{equation}
  \nu \equiv \left( {\delta_c(z)\over \sigma(M)} \right)^2,
\end{equation}
where $\delta_c(z)$ is the spherical over-density which collapses at $z$ ($\delta_c \approx 1.68$)  
and $\sigma(M)$ is the rms fluctuation in the matter density smoothed on a scale $R=(3M/4\pi \bar{\rho})^{1/3}$.
$\nu(M_*) = 1$ defines the non-linear mass scale.   A cosmology with $\Omega_0 = 0.3$, $\Lambda = 0.7$, $\Gamma=0.2$, and $\sigma_8=0.9$ has $M_* \approx 10^{13}\ h^{-1}\ M_{\sun}$. 

The power spectrum has two terms.  The first corresponds to correlations in density between pairs of points where each member of the pair lies in a different halo, and so is named the ``halo-halo'' (hh) term.  The second corresponds to correlations between pairs in the same halo, and is known as the one-halo or 
Poisson (P) term.  For convenience in calculating convolutions, we work in Fourier space, introducing the Fourier transform of the halo profile, normalized by the virial mass, 
\begin{equation}
y(M,k)={1 \over M} \int 4 \pi r^2\ \rho(r) {\sin (kr) \over kr}\, dr. 
\label{eqn:y}
\end{equation}

The mass of the NFW profile is logarithmically divergent, so 
in order to evaluate this integral, we must impose a cutoff. This does not have to be at the 
virial radius, since we know that haloes are not completely truncated there. 
Instead, NFW profile typically continues to $2-3r_{\rm vir}$. In this 
regime there is already some overlap between the haloes, so for the purpose 
of correlations one 
can count the same mass element in more than one halo. Thus the mass function 
integrated over all the haloes may even exceed the mean density of the universe 
(but it can also be below it, since it is not required that all the matter 
should be inside a halo). 

The halo-halo contribution is \begin{equation}
P^{hh}(k) = P_{\rm lin}(k) \left[
  \int d\nu f(\nu) d\nu\ b(\nu) y(M(\nu),k) \right]^2,
\label{eqn:twohalo}
\end{equation}
where $b(\nu)$ is the (linear) bias of a halo of mass $M(\nu)$, for which we use 
the $N$-body fit of \citet{1999MNRAS.308..119S}. Since we want this term to reproduce the linear power 
spectrum on large scales we impose this constraint onto the form for $b(\nu)$ 
\citep{2000MNRAS.318..203S}.
The Poisson contribution is 
\begin{equation}
  P^{P}(k) = {1\over (2\pi)^3} \int d\nu f(\nu)
  \ {M(\nu)\over\bar{\rho}} |y(M(\nu),k)|^2.
\label{eqn:Poisson}
\end{equation}
If a cutoff beyond the virial radius is used, the mass weighting should reflect the increase.
The treatment of the Poisson term on large scales is only approximate. 
Mass and momentum conservation require that on very large scales the
non-linear term 
should scale as $k^4$, rather than as a constant implied by equation \ref{eqn:Poisson}, 
so the contribution from this term on large scales is overestimated for $k\ll r_{\rm vir}^{-1}$. 
This partially compensates the increase in power from the matter outside the virial 
radius and for this reason we chose to use $r_{\rm vir}$ as the radial cutoff for the
halo. The halo-halo term is only approximate, since we do not include the 
exclusion of haloes, which would suppress the term. 
Because of these approximations one would not expect the 
halo model to be perfect, especially in the transition between the linear and non-linear 
regime.

The total power spectrum is the sum of the two contributions,
$P(k)=P^{hh}(k)+P^P(k)$.
On small scales, the Poisson term dominates.  On larger scales, the halo-halo term dominates and reduces to the linear power spectrum.  From here forward, we express the power spectrum in dimensionless form, $\Delta^2(k)=4\pi k^3 P(k)$.

\begin{figure}
\begin{center}
\includegraphics[scale=0.95]{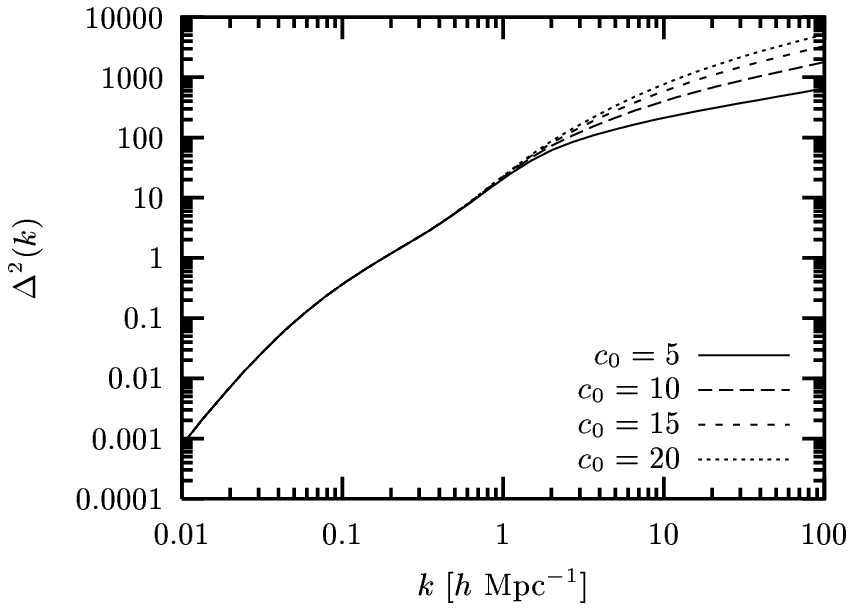}
\\
\includegraphics[scale=0.95]{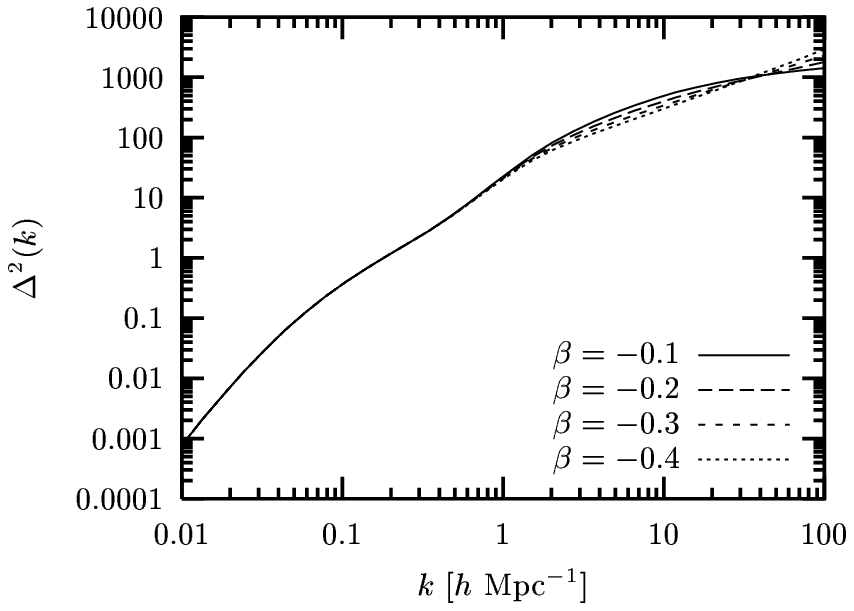}
\end{center}
\caption{Using the halo model, several power spectra were generated from a cosmology with $\Omega_0 = 0.3$, $\Lambda = 0.7$, $\Gamma=0.2$, and $\sigma_8=0.9$ at $z=0$.  In the upper panel, $c_0$ varies and $\beta=-0.2$.  In the lower panel, $\beta$ varies and $c_0=10$.}
\label{fig:varyc0beta}
\end{figure}

The effect on the power spectrum of varying $c_0$ and $\beta$ in a typical $\Lambda$CDM cosmology is shown in Figure \ref{fig:varyc0beta}.
The concentration controls how tightly matter is correlated within a single halo.  Therefore, higher concentration means a larger one-halo term.  
Increasing $c_0$ increases the concentration at the non-linear mass, affecting
the amplitude of the power spectrum where it is dominated by one-halo term.  

Varying $\beta$ keeping $c_0$ constant produces a tilt in the non-linear 
power spectrum around the fiducial value, which is set by the scale where the
non-linear mass dominates the power spectrum contribution ($k \sim 30$--$40$ $h$ Mpc$^{-1}$).
Haloes less massive than the non-linear mass dominate at higher $k$.  Steeper (more negative) $\beta$ 
means that haloes less massive than the non-linear mass will have their concentrations enhanced, leading to an enhanced one-halo
term, and more power at high $k$.  Haloes more massive than the non-linear mass will have their concentration reduced.  These dominate at intermediate $k$ in the non-linear regime, so power there is reduced.  The changes in $\beta$ shown in the figure may not have much effect on the power spectrum at $k<100$ $h$ Mpc$^{-1}$, which is the range where $N$-body simulations are reliable. 
On the other hand, they
have a substantial effect on the concentration of typical-sized haloes.  
For the same concentration at the non-linear mass ($\sim 10^{13}\ h^{-1}\ M_{\sun}$), a halo at $10^{12}$ $h^{-1}$ $M_{\sun}$ has a concentration 50\% lower with $\beta=-0.1$ than with $\beta=-0.4$.  These masses do not dominate the power spectrum until $k>100$ $h$ Mpc$^{-1}$, 
so this change does not make as much of a difference on the power spectrum at scales larger than this.
This discussion suggests that while power spectrum analysis cannot provide 
strong constraints on the mass dependence of concentration, one can use 
concentration mass relation to predict the power spectrum at small scales
which are not resolved by simulations. 

\section{Method and Results} \label{sec:Method}
We consider two sets of cases which have been extensively simulated, 
self-similar initial conditions and  
more realistic cold dark matter initial conditions. 
In both cases we use the fitting formulae for non-linear $\Delta^2(k)$ given by 
\citet{2002astro.ph..7664S}.

\subsection{Self-Similar Case} \label{subsec:Powerlaw}
The simplest case to consider is 
$\Omega=1$ Einstein-de Sitter universe with a power law linear power 
spectrum.  In this case, $\beta$ 
has an analytic form, provided we assume that once a halo collapses, the scale 
radius $r_s$ is fixed in proper coordinates. This is suggested by the simple 
model which assumes the haloes remain unchanged once formed,  
which seems to hold in numerical simulations \citep{2001MNRAS.321..559B}.  
We follow the evolution of a single halo, as it traces out a portion of the $c(M)$ relation.

In an $\Omega=1$ universe, density perturbations grow as the scale factor $a=1/(1+z)$.  For a power law linear spectrum, this means $P_{\rm lin}(k) \propto a^2 k^n$.  Top-hat smoothing at the scale corresponding to $M$ yields 
$\sigma(M) \propto a M^{-{n+3 \over 6}}$. 
For an $\Omega_m=1$ universe, $\delta_c \approx 1.68$ is constant in time, so  $\sigma(M_*)$ is also constant.  This means
$M_*^{n+3 \over 6} \propto  a$.
If $r_s$ is fixed in time, $c \propto r_{\rm vir} \propto a$ 
follows from the definitions of $c$ and $r_{\rm vir}$ and from $\bar\rho \propto a^{-3}$.  
Since $c \propto M_*^{-\beta}$ then $ 
\beta \approx -{n+3 \over 6}$ if we assume $M$ is constant. In reality $M$ also increases with $a$, which 
decreases $\beta$ somewhat, although not by more than 20\%. 

\begin{figure}
\begin{center}
\includegraphics[scale=0.95]{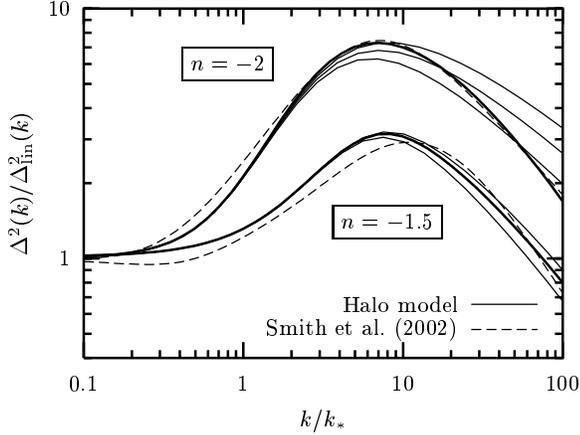}
\end{center}
\caption{Non-linear power spectra in scale free models, divided by the linear power.  The power law linear spectra are $\Delta^2_{\rm lin}=4 \pi k^3 (k/k_*)^n$, where $n=-2$ and $n=-1.5$.  In each case the halo model spectra for several different $c(M)$ are shown in solid lines, and the fitting formula of Smith et al. (2002) is in dashed lines.  For the $n=-2$ case we show three models with $\beta=.-1667$, which is the analytically predicted value from the text.  These have $c_0=3,4,5$ in increasing amplitude, and fit poorly.  We also show the best-fitting $\beta=-.037$, $c_0=2.6$ with a thicker line.  For $n=-1.5$, we show the predicted $\beta=-.25$ with $c_0=6,8$, and the best-fitting $\beta=-.245$, $c_0=7.1$.}
\label{fig:scalefree}
\end{figure}

We calculated halo model power spectra with $n=-2.0$ and $n=-1.5$, for which $\beta=-0.16$ and $-0.25$ as calculated above, with several values of $c_0$.  These spectra are shown 
in Figure \ref{fig:scalefree}. The agreement is quite good, given all the limitations of the halo model. 
At higher $k$ in the $n=-2.0$ case, it appears that the slope of the power spectrum disagrees independent of 
$c_0$. 
For $k/k_*>2$ this model is better fit by a power spectrum with $\beta=-.037$, $c_0=2.6$. 
However, 
according to figure 12 of \citet{2002astro.ph..7664S}, simulation data do not exist for $k/k_*$ greater than a few tens in the $n=-2$ case, 
so the discrepancy is not really significant. 
In the $n=-1.5$ case, which is tested against simulations to $k/k_*=100$, the predicted $\beta=-0.25$ is close 
to the best fitted value. This agreement 
therefore confirms the assumption that haloes have a fixed 
scale radius once they formed.

\subsection{Cold Dark Matter Models} \label{subsec:Results}

Next we consider several flat CDM models (we do not consider open or closed models
here, since they are observationally disfavored). 
To compare the halo model power spectrum to the  \citet{2002astro.ph..7664S} 
power spectrum, we use a simple 
$\chi^2$ statistic:
\begin{equation}
\chi^2 = \sum^N_{i=1} \left( {\Delta^2_{\rm Smith}(k_i)-\Delta^2(k_i) \over \sigma_i}\right)^2.
\end{equation}
We distributed the $N=40$ sample $k_i$ evenly in $\log k$, considering $0.01<k< 40$ $h$ Mpc$^{-1}$.  This range takes us from the linear regime to some of the highest $k$ that the fitting formula has been tested with, according to figure 16 of \citet{2002astro.ph..7664S}.  For the errors $\sigma_i$, we take 30\% of the \citet{2002astro.ph..7664S} power.  This is somewhat arbitrary, but roughly the size of the combined error 
in the mildly non-linear regime, and is probably conservatively large in the fully non-linear regime where the Poisson term dominates. 
We are ignoring the correlations between the points, 
so the actual value of $\chi^2$ is just a qualitative measure of the goodness of fit. 

For a variety of cosmologies, we minimized $\chi^2$ over the two-dimensional parameter space of $c_0$ and $\beta$ from equation (\ref{eqn:c(M)}).  We employed a Powell minimization as described in \citet{1992nrfa.book.....P} to give a best-fitting $c_0$ and $\beta$.  All runs were given the same initial point: $c_0=10$ and $\beta=-0.2$, and terminated when the the minimum changed by less than 0.1\%.  We place an error bar on each minimization by calculating $\chi^2$ on a grid about the minimum point.  We compute an error contour where $\exp(-\chi^2/2)$ falls to half-maximum.  The error bars in $c_0$ and $\beta$ show the maximum extent of this contour in each direction.  With this definition, these error bars cannot properly be used to set confidence limits or rule out models, but are included simply to gauge in which cosmologies parameters are better or worse constrained and to explore parameter degeneracies.  Indeed, because of the range of $k$ we consider, $c_0$ and $\beta$ are somewhat degenerate: making $\beta$ more negative has a similar effect to lowering $c_0$ (compare Figure \ref{fig:varyc0beta}).

\begin{figure}
\begin{center}
\includegraphics[scale=0.95]{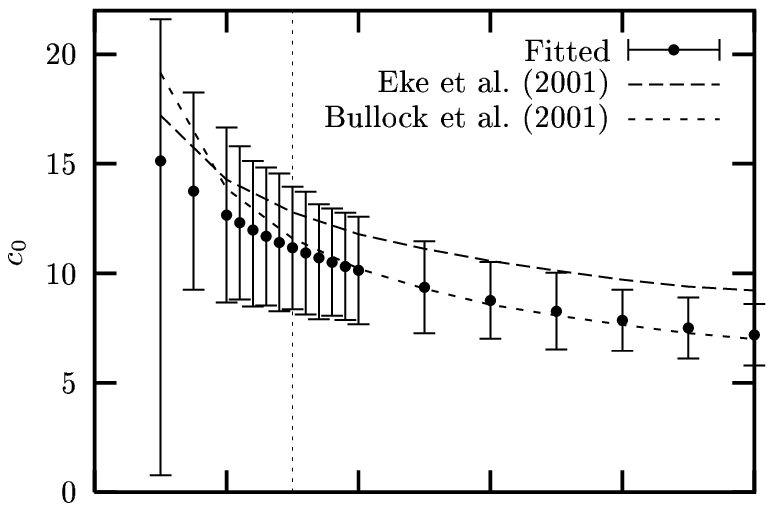}
\\
\includegraphics[scale=0.95]{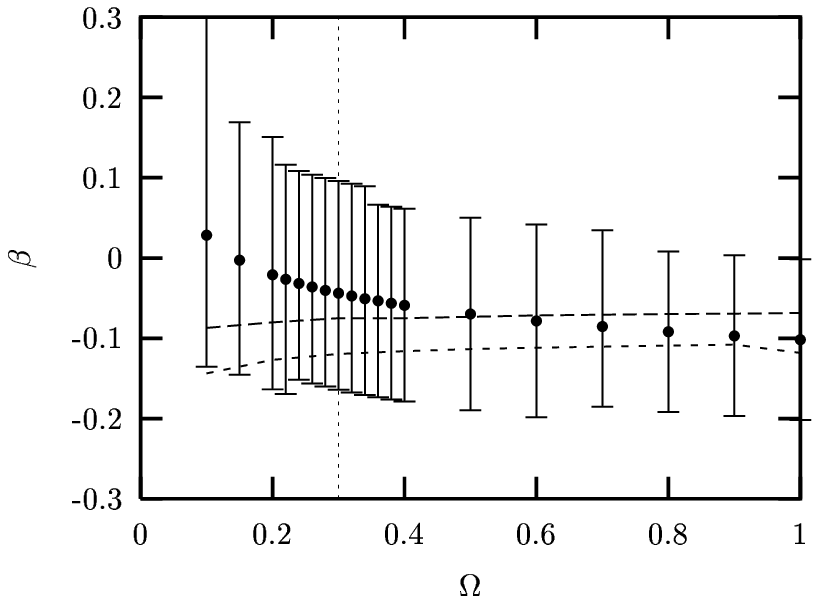}
\end{center}
\caption{The values of $c_0$ and $\beta$ at several values of the mass density parameter $\Omega_0$ for the fitted halo model power spectrum and the models compared in \S\ref{sec:Comparison}.  The vertical line denotes the fiducial model with $\Omega_0 = 0.3$.  The other cosmological parameters are given in \S\ref{subsec:Results}.}
\label{fig:varyOmega}
\end{figure}

We varied one cosmological parameter at a time, choosing the other parameters from a fiducial model with $\Omega_0 = 0.3$, $\Omega_0 + \Lambda = 1$, $\Gamma=0.2$, $\sigma_8=0.9$, and $z=0$.  We examined the ranges $0.1<\Omega_0<1.0$, $0.1<\Gamma<0.4$, $0.6<\sigma_8<1.5$,
and $0<z<1$.
We plot the variation of $c_0$ and $\beta$ with $\Omega_0$ with solid lines in Figure \ref{fig:varyOmega}.  The dashed lines are discussed in \S\ref{sec:Comparison}.
The relation for $c_0$ is well fit by:
\begin{equation}
c_0(\Omega_0)	=	11 (\Omega_0/0.3)^{-0.35},
\end{equation}
while $\beta$ is consistent with being constant around zero or slightly negative.
Note that the $\Omega$ dependence depends on the definition of the 
virial radius and with a different definition there would be a different
$\Omega$ dependence. 

\begin{figure}
\begin{center}
\includegraphics[scale=0.95]{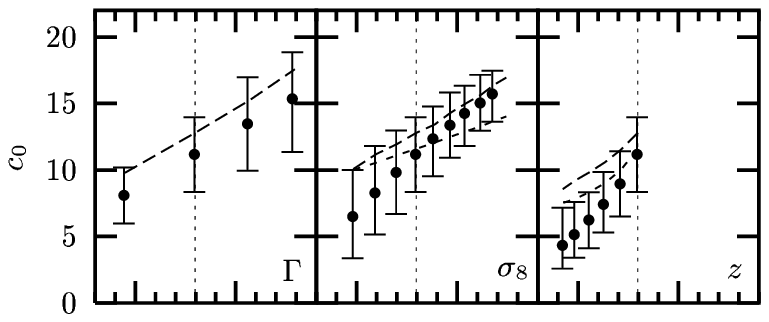}
\\
\includegraphics[scale=0.95]{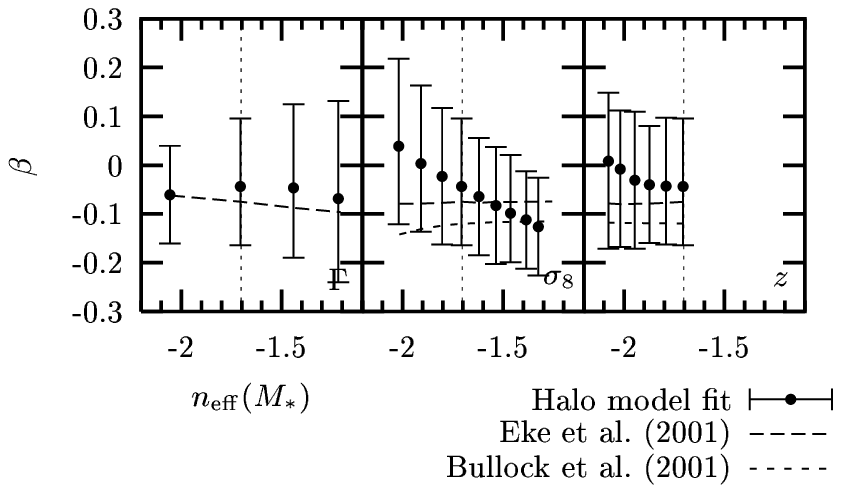}
\end{center}
\caption{The values of $c_0$ and $\beta$ as a function of the effective spectral index at the non-linear scale $n_{\rm eff}$ for the fitted halo model power spectrum and for the models compared in \S\ref{sec:Comparison}.  The line type indicates which model.  We vary one parameter at a time.  The fiducial model, given in \S\ref{subsec:Results}, is marked the vertical line at $n_{\rm eff} \approx -1.7$.  Each parameter increases as $n_{\rm eff}$ increases except for $z$, which decreases.  In their respective panels, the points (left to right) correspond to: $\Gamma = 0.1,0.2,0.3,0.4$, $\sigma_8 = 0.6, 0.7, 0.8, 0.9, 1.0, 1.1, 1.2, 1.3, 1.4$, and $z= 1.0,0.8,0.6,0.4,0.2,0.0$.}
\label{fig:neff}
\end{figure}

Although $c_0$ and $\beta$ vary with all of
the other cosmological parameters, it is more instructive to present its
variation with the
effective power law index of the linear power spectrum at the non-linear scale, 
\begin{equation}
n_{\rm eff}=\left. {d(\log P_{\rm lin}(k))\over d(\log k)} \right| _{k_*},
\end{equation}
where $\Delta^2_{\rm lin}(k_*)=1$ defines the non-linear scale. The reason is 
that the slope of the linear power spectrum determines the epoch of 
formation of small haloes that merge into the larger halo.
For lower effective slope $n_{\rm eff}$ these haloes formed later, when 
the density of the universe was smaller, and the final halo concentration of
the larger halo is also lower. 
This is depicted with solid lines in Figure \ref{fig:neff}.    
Our fiducial cosmology has $n_{\rm eff} \approx -1.7$.  Varying $\Omega_0$ has no effect on $n_{\rm eff}$.  When varying $z$, a portion of the variation of $c_0$ and $\beta$ should be due to the change in $\Omega(z)$.  
The variation of $c_0$ with $n_{\rm eff}$ is consistent with
\begin{equation}
c_0 = 11  (\Omega(z)/0.3)^{-0.35} \left({n_{\rm eff}\over -1.7}\right)^{-1.6},
\end{equation}
while $\beta$ has large errors, but is typically slightly negative.

\section{Comparison to halo analysis} \label{sec:Comparison}
We now compare our results
to the models of \citet{2001ApJ...554..114E} and \citet{2001MNRAS.321..559B}. 
These are designed to reproduce the concentrations of haloes found in $N$-body simulations.  
Throughout we translate to our virial convention.  

\begin{figure}
\begin{center}
\includegraphics[scale=.95]{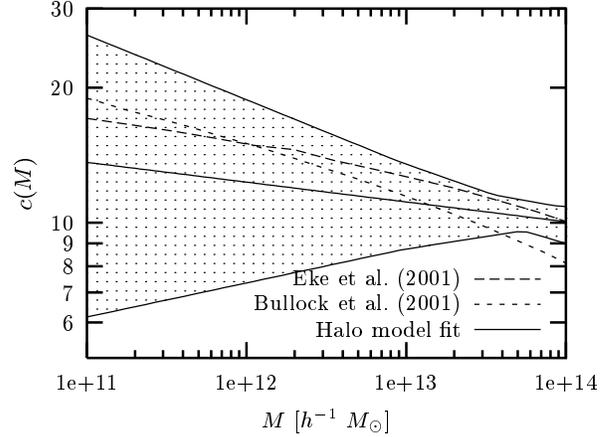}
\end{center}
\caption{A comparison of the variation of the concentration parameter with mass for our fiducial $\Lambda$CDM cosmology (\S\ref{subsec:Results}).  The masses and concentrations here have been translated to a common virial radius convention, discussed in the text.  The shaded region covers power law models within our error contour, discussed in the text, and should not be considered a limit on $c(M)$ relations of arbitrary shape.}
\label{fig:c(M)}
\end{figure}

We begin by considering concentration as a function of halo mass for 
the fiducial $\Lambda$CDM cosmology defined in \S\ref{subsec:Results}.  This is plotted in Figure \ref{fig:c(M)}.  
It is clear that our results are in agreement with those from direct 
halo structure analysis. 
However, most of the information in our model comes from the haloes with masses between $10^{13}$ and 
$10^{14}\ h^{-1}\ M_{\sun}$, because haloes smaller than this
dominate at scales smaller than the range we considered for our fit.  
This means we have a relatively short lever arm with which to determine the mass concentration relation and consequently 
we are unable to strongly constrain the slope $\beta$. 
In direct analysis of haloes, concentration is a decreasing function of mass, so $\beta<0$.  
Our best fitted values for $\beta$ are also typically negative, although the errors 
are large and positive values also cannot be excluded. The strongest confirmation of 
this prediction comes from the analysis of scale-free models in previous section, where 
$\beta$ is negative for $n=-1.5$. 
It is clear that measuring the mass-concentration directly is a better 
way to determine the mass dependence of concentration rather than using the 
power spectrum, particularly at lower masses.

\begin{figure}
\begin{center}
\includegraphics[scale=0.95]{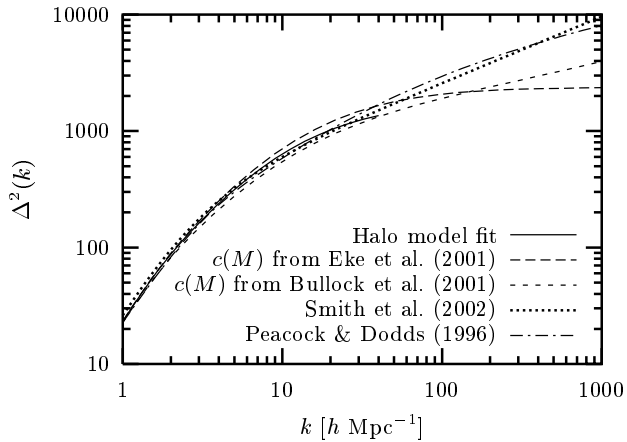}
\end{center}
\caption{A comparison of power spectra generated with different $c(M)$ with analytic fitting formulae.  Note that we only fit up to $k=40$ $h$ Mpc$^{-1}$.}
\label{fig:Delta2}
\end{figure}

Since determining mass dependence of concentration from the power spectrum does 
not appear promising let us ask the opposite question: how well can we predict 
the power spectrum using the mass concentration relations measured directly 
from the analysis of individual haloes in the simulations? 
Figure \ref{fig:Delta2} shows the power spectra calculated with power law approximations 
to the $c(M)$ functions shown in Figure \ref{fig:c(M)}, as well as the \citet{2002astro.ph..7664S} 
power spectrum. While there is good agrement between them for $k<40h$Mpc$^{-1}$, 
the halo models always predicts less power at $k>40h$Mpc$^{-1}$ compared to the fits
to the power spectrum. However, the direct fits are not reliable in this regime, since 
the $N$-body simulations become unreliable for  $k>40h$Mpc$^{-1}$, which is 
why the fits presented above only use information from $k<40h$Mpc$^{-1}$. 

The conclusion from the above comparison is that on small scales 
there is disagreement between 
analytic formulae for non-linear power spectrum and the halo model. 
Since the analytic models have not been calibrated with simulations 
in this regime it is possible that they are unreliable and so 
the halo model should be used instead (indeed, the analytic models of
\citealt{2002astro.ph..7664S} do not even predict the power spectrum for $k>100h$Mpc$^{-1}$). 
Nevertheless, since non-linear predictions at high $k$ have been used 
in the literature, especially for predicting the lensing effects on 
small scales (e.g. \citealt{1999MNRAS.305..746M}), it is important to recognize their possible 
limitations when using them outside the regime of applicability. 

It is unlikely that the discrepancy can be resolved by modifying the 
mass function. The existing simulations resolve the mass function
down to $10^{11}M_{\sun}$, which is the mass that dominates the
power spectrum at $k\sim 1000h$Mpc$^{-1}$. Substructure could 
boost the amount of power, since in the mass function one counts only 
isolated haloes, not those that are within larger haloes. Subhaloes
within haloes contribute to the clustering in the same way as
isolated haloes on scales that are comparable to the scale radius.  
However, the abundance of subhaloes is small compared to the 
abundance of isolated haloes even at small halo masses, so it is 
unlikely that this is a significant correction \citep{2001MNRAS.321..559B}. 
Similarly, while the scatter in the mass concentration relation 
boosts the amount of power, we find the effect is relatively small
for the scatter suggested from simulations \citep{2001MNRAS.321..559B,2001ApJ...554...56C} and again 
cannot resolve the discrepancy. 

For $c_0$ versus $\Omega_0$ and  $n_{\rm eff}$ our fit to \citet{2002astro.ph..7664S} 
produces very similar trends to that in \citet{2001ApJ...554..114E} and \citet{2001MNRAS.321..559B}. 
This comparison should be reliable, since the haloes that dominate are around 
$10^{13}-10^{14}M_{\sun}$, which are abundant and well resolved in the simulations. 
For $\beta$ we also find broad agreement, although the errors from the power spectrum 
method are very large.

\section{Conclusions} \label{sec:Conclusions}
We have used the halo model and the non-linear evolution of dark matter power 
spectrum (as given by the fitting formula of \citealt{2002astro.ph..7664S}) 
to probe the dependence of halo concentration on cosmological parameters.  The most important parameters for the concentration are the matter density and the effective slope
of the linear power spectrum at the non-linear scale. 
We present analytical expressions 
which give concetration at the non-linear scale as a function of these two parameters.

We examined our result against the models for halo concentration of \citet{2001ApJ...554..114E} and  
\citet{2001MNRAS.321..559B}. We found broad agreement with the trends presented there in 
that we also find that spectra with lower effective slope $n_{\rm eff}$ 
have lower concentration at the non-linear mass. 
Although there is some difference in the mass-concentration dependence 
between \citet{2001ApJ...554..114E} and
\citet{2001MNRAS.321..559B},
we find
no compelling reason to prefer one model to the other based on the power spectrum.
This is because the dynamic range covered by the power spectrum analysis is too narrow to obtain 
a reliable mass dependence of the concentration over a wide range of masses. We have found that 
using these models to predict the non-linear power spectrum results in a significantly lower 
power spectrum on small scales compared to \citet{1996MNRAS.280L..19P} and 
\citet{2002astro.ph..7664S}. This 
suggests that care must be exercised when using these models in the range where they have 
not been calibrated and may indicate that the amount of dark matter power on small scales 
is smaller than predictions from these models. 

For mildly non-linear $k$, even the best fitted halo models show discrepancies from 
the \citet{2002astro.ph..7664S} power spectrum
at the 20-30\% level, both for 
the self-similar and CDM models. Some of this discrepancy arises from 
the fits in \citet{2002astro.ph..7664S}, which 
are only accurate at 10\% level. However, direct comparison of the halo model to the 
$N$-body simulations of \citet{1998ApJ...499...20J} also reveals discrepancies at the similar level.   
This is not surprising given the approximate nature of the 
halo model and argues that it
cannot be a full replacement for $N$-body simulations in the 
era of high precision cosmology. However, using the parameters derived 
in this paper, the halo model can be used 
for qualitative predictions of the dark matter power 
spectrum over much wider range of models and scales than previously 
available. 

\section*{Acknowledgments}
We thank J.~Bullock, V.~Eke, J.~A.~Peacock and R.~E.~Smith for making their codes available. 
KH is supported by an NSF Graduate Research Fellowship.
US is supported by NASA, NSF, Sloan and Packard Foundations. 

\bibliography{cosmo,cosmo_preprints}\bibliographystyle{mnras}

\end{document}